# POWER-SPECTRUM ANALYSES OF SUPER-KAMIOKANDE SOLAR NEUTRINO DATA: VARIABILITY AND ITS IMPLICATIONS FOR SOLAR PHYSICS AND NEUTRINO PHYSICS

P.A. Sturrock[1], D.O. Caldwell[2], J. D. Scargle[3], and M.S. Wheatland[4]

[1] Center for Space Science and Astrophysics, Varian 302, Stanford University, Stanford, California 94305-4060
[2] Physics Department, University of California, Santa Barbara, CA 93106-9530
[3] NASA/Ames Research Center, MS 245-3, Moffett Field, CA 94035
[4] School of Physics, University of Sydney, Sydney, Australia

Since rotational or similar modulation of the solar neutrino flux would seem to be incompatible with the currently accepted theoretical interpretation of the solar neutrino deficit, it is important to determine whether or not such modulation occurs. There have been conflicting claims as to whether or not power-spectrum analysis of the Super-Kamiokande solar neutrino data yields evidence of variability. Comparison of these claims is complicated by the fact that the relevant articles may use different datasets, different methods of analysis, and different procedures for significance estimation. The purpose of this article is to clarify the role of power spectrum analysis. To this end, we analyze primarily the Super-Kamiokande 5-day dataset, and we use a standard procedure for significance estimation as used by the Super-Kamiokande collaboration. We then analyze this dataset, with this method of significance estimation, using six methods of power spectrum analysis. Five of these have been used in published articles, and the other is a method that might have been used. We find that, with one exception, the results of these calculations are consistent with those of previously published analyses. We find that the power of the principal modulation (that at 9.43 yr$^{-1}$) is greater in analyses that take account of error estimates than in the basic Lomb-Scargle analysis that does not take account of error estimates. The corresponding significance level reaches 99.3% for one method of analysis. However, we find a problem with the recent article by Koshio: we can reproduce the results of his power-spectrum analysis, but not the results of his Monte-Carlo simulations. We have a suggestion that may account for the difference. We also comment on a recent article by Yoo et al. We discuss, in terms of subdominant processes, possible neutrino-physics interpretations of the apparent variability of the Super-Kamiokande measurements, and we suggest steps that could be taken to resolve the question of variability of the solar neutrino flux.

1. INTRODUCTION

Combined evidence from many experiments shows that the solar-neutrino deficit may be attributed to matter-induced oscillations within the Sun - specifically, the large-mixing angle version of the MSW process [1]. This model incorporates the assumption that, since the density profile of the Sun is highly stable, the intrinsic solar neutrino flux should be constant. If it is found that the solar neutrino flux is intrinsically variable, we must conclude that either our current understanding of the Sun or our current understanding of the neutrino model is inadequate. Possible interpretations in terms of solar physics would include non-steady or non-spherically-symmetric nuclear burning. A time variation with properties that point to association with the solar magnetic field would comprise evidence that the explanation is to be found in a more complex neutrino mechanism.

Current observational and experimental evidence does not preclude the possibility that the solar neutrino flux may be influenced by (in addition to the dominant MSW effect) a subdominant process that may involve both flavor change and spin change. While a Resonant-Spin-Flavor-Precession (RSFP) process [2] itself gives an excellent fit to solar neutrino data, it must, if it occurs at all, be subdominant (subordinate) to the MSW effect because of the observation of a decrement in both solar $\nu_e$ flux in a strong magnetic field and reactor $\bar{\nu}_e$ flux [3] in no such field. Should the RSFP process involve only the three well-established active neutrinos, it can take place only in the solar core [4], because of the known neutrino mass differences.

Another possibility [5] is that the solar electron neutrino may convert to a sterile neutrino via a transition magnetic moment. Analysis of this form of the RSFP process shows that it can provide a better fit to time-averaged solar neutrino measurements than is provided by the MSW process alone [6]. Since this sterile neutrino does not mix with active neutrinos, all known constraints on the sterile neutrino become irrelevant. Evidence for a sterile neutrino would provide a window on new physics that may prove to be more important than the discovery of neutrino oscillations.

For these and other reasons, it is important to determine whether or not the solar neutrino flux varies with time and, if so, whether there is evidence that the modulation is related to the solar magnetic field. It is advantageous that the rotation period of the Sun (about 27 days as seen from Earth) is short enough that the powerful procedures of power spectrum analysis are applicable. The most extensive available solar neutrino dataset is that produced by the Super-Kamiokande collaboration [7]. It must be emphasized, however, that the subdominant flux modulation would be quite small. For the Super-Kamiokande energy region, the expected modulation is of order 2% for the three-neutrino model [4], and rather larger for the four-neutrino case, for which the solar electron neutrino survival probability is a fairly flat function of energy in this range. It is therefore necessary to use power-spectrum analysis procedures that can detect small depths of modulation.

There have been a number of recent articles presenting power-spectrum analyses of the Super-Kamiokande solar neutrino data. Those published by members of the Super-Kamiokande collaboration claim that there is no evidence for variability [7 - 9]. On the other hand, investigators outside the collaboration have claimed to find evidence for variability. Milsztajn [10] claimed to find evidence for variability in an analysis of Super-Kamiokande 10-day data, and we have presented evidence in favor of variability from analyses of both the 10-day data [11 - 13] and 5-day data [5, 14]. Early articles worked with the first dataset, which was organized in 10-day bins, but later analyses used the second dataset, which was organized in 5-day bins. However, the methodologies have not been uniform (there are differences in significance estimation, search bands, etc.), so it is important to understand the extent to which the differences in the claimed conclusions may be attributed to methodological differences.

For this reason, we have carried out a sequence of power-spectrum analyses of the Super-Kamiokande 5-day dataset, using in turn the five methods that have been used so far, and another that might have been used. We standardize other factors by (a) concentrating on the 5-day dataset, (b) adopting a standard search band (0 to 50 yr$^{-1}$), and (c) using Monte-Carlo simulations for significance estimation. The 5-day dataset [7] lists the start times, end times, mean live times, flux estimates, lower error estimates, and upper error estimates for each of 358 bins, beginning on May 31, 1996, and ending on July 15, 2001. We over-sample the power spectrum by adopting a frequency resolution of 0.01 $yr^{-1}$ in order to obtain reliable estimates of the peak values of the power.

In Section 2, we give the results of a simple Lomb-Scargle analysis [15, 16] of the 5-day dataset, using only the mean times and the flux estimates, as in the Milsztajn [10] analysis of the Super-Kamiokande 10-day data, and (following Yoo et al. [7]) standardizing the frequency interval to be 0 to 50 $yr^{-1}$. We also repeat the Lomb-Scargle analysis, incorporating the mean live times rather than the mean times, as in the analyses of Nakahata [8] and Yoo et al. [7]. This results in little change in the power spectrum. We also carry out spectrum analysis by a method, based on a proposal by Scargle [17], which takes account of the experimental error estimates. This leads to a notable change in the power spectrum, that now has a pronounced peak at 9.43 $yr^{-1}$ with power 9.56.

For each analysis, we present the results of Monte-Carlo simulations based on the assumption that there is no real modulation in the time series (a "false-alarm" analysis). We also carry out an analysis relevant to the sensitivity of the Lomb-Scargle procedure: If the modulation at 9.43 $yr^{-1}$ were real, what power should we expect to find at that frequency? The results are somewhat surprising.

In Section 3, we carry out spectrum analyses by likelihood methods [18]. The first analysis takes account of the error estimates and of the start time and end time. The leading peak remains that at 9.43 $yr^{-1}$. We next present the results of an analysis that takes account also of the mean live time. This modification makes little change in the resulting power spectrum. We also present the results of an analysis that allows for a "floating offset," as used in our analyses of Homestake [18] and GALLEX [19] data, and as used recently by Koshio [9]. The results differ little from those of the preceding likelihood calculations.

For each of these methods, we present the results of Monte-Carlo simulations. These calculations bear on the possibility of a "Type 1" (or "false alarm") error: What is the probability that what we consider to be evidence for modulation has arisen purely by chance?

For the floating-offset method, as for the Lomb-Scargle method, we also present calculations relevant to the sensitivity of the method. This calculation bears on the possibility of a "Type 2" (or "missed conflagration") error: If the modulation is real, what is the probability that an experiment similar to, but not identical to, the Super-Kamiokande experiment would *fail* to detect it? As for the Lomb-Scargle case, the results of this calculation are rather surprising.

In Section 4, we carry out a comparative analysis of the 10-day and 5-day datasets, focusing on the role of aliasing. Since each dataset was sampled in a highly regular time sequence, modulation at a given frequency can appear, in a power spectrum, not only at that frequency but also at one or more related frequencies. There is unmistakable aliasing of the primary modulation at 9.43 yr$^{-1}$. However, we also find evidence for aliasing of a peak at 12.31 yr$^{-1}$, which is an interesting frequency since it may be related to solar rotation.

In Section 5, we review the results of the previous sections, comparing the results with those of other authors. We find that our results are quite consistent with results previously published by Milsztajn [10], Nakahata [8], and Yoo et al. [7]. We find an inconsistency between our results and those of Koshio [9], and we speculate on a possible cause of the discrepancy. We also review briefly the significance of possible time variation of the solar neutrino flux for neutrino physics and for solar physics.

## 2. LOMB- SCARGLE ANALYSES

The Super-Kamiokande 5-day data are organized in bins which we enumerate by $r = 1,\ldots,R$, where $R = 358$. For each bin we are given the start time $t_{s,r}$, the end time, $t_{e,r}$, the "weighted mean live time" $t_{ml,r}$, the flux estimate $g_r$, the lower error estimate $s_{l,r}$, and the upper error estimate $s_{u,r}$. We find that the two error estimates have a close relationship: the ratio of the upper error estimate to the lower error estimate has a mean value of 1.17 with a standard deviation of only 0.046. For this reason, we here work with a single error estimate formed from their mean:

$$s_r = \tfrac{1}{2}(s_{l,r} + s_{u,r}). \tag{2.1}$$

We now normalize the flux estimates

$$x_r = \frac{g_r}{mean(g_s)} - 1, \qquad (2.2)$$

and also the error estimates

$$\sigma_r = \frac{s_r}{mean(g_s)}. \qquad (2.3)$$

However, the experimental error estimates are not used in the basic Lomb-Scargle calculations of this section.

Following Lomb [15] and Scargle [16] (see also Press et al. [20]), we form a power spectrum from

$$S(\nu) = \frac{1}{2\sigma_0^2} \left\{ \frac{\left[\sum_r x_r \cos(2\pi\nu(t_r - \tau))\right]^2}{\sum_r \cos^2(2\pi\nu(t_r - \tau))} + \frac{\left[\sum_r x_r \sin(2\pi\nu(t_r - \tau))\right]^2}{\sum_r \sin^2(2\pi\nu(t_r - \tau))} \right\}, \qquad (2.4)$$

where

$$\sigma_0 = std(x_r), \qquad (2.5)$$

and $\tau$ is defined by the relation

$$\tan(4\pi\nu\tau) = \frac{\sum_r \sin(4\pi\nu t_r)}{\sum_r \cos(4\pi\nu t_r)}. \qquad (2.6)$$

In order to use the Lomb-Scargle procedure, it is necessary to assign a definite time $t_r$ to each bin. In this section, we adopt the mean of the start and end time, as in the early work of Milsztajn [10]. This yields the power spectrum shown in Figure 1. The top ten peaks are listed in Table 1. Of the five leading peaks, those at frequencies 9.43 yr$^{-1}$, 12.31 yr$^{-1}$, 39.28 yr$^{-1}$, and 43.72 yr$^{-1}$ recur in later analyses.

Here and in later sections, we assess the significance of the leading peak by Monte-Carlo simulations. We generate a large number of simulated datasets by the algorithm

$$x_{MC,r} = \sigma_r randn, \qquad (2.7)$$

where randn is the operation of producing random numbers with a normal distribution, zero mean, and variance unity. (Yoo et al. [7] use a fixed value of the error estimate [$\sigma_0$] for all data points, while we use the value (2.3) based on the error estimates given by the experimenters for each data point.) For each fictitious dataset, we compute the power spectrum over the range 0 to 50 yr$^{-1}$, and note the power of the highest peak, which we denote by SM for "spectral maximum." We then examine the distribution of the maximum-power values.

We present the results of these simulations in Figure 2, which shows the distribution of values of SM from the simulations, and indicates the value of SM for the actual data. We see that 49% of the simulations have power equal to or exceeding that of the strongest peak (S = 6.79 at frequency 43.72 yr$^{-1}$) in the actual power spectrum. We also find that 824 out of 1,000 simulations have power larger than the power (5.90) at frequency 9.43 yr$^{-1}$, which will prove to be the frequency of most interest in our later likelihood analyses. From this perspective, one would conclude that there is no evidence for a periodic modulation of the neutrino flux. It is important to note that this test assumes that, a priori, all frequencies in the chosen search band (here 0 to 50 yr$^{-1}$) are equally likely. Hence we are ignoring all available information concerning variability in solar structure and dynamics. We comment on this point further in Section 5.

The Lomb-Scargle procedure is equivalent to finding a least-squares fit of a sine-wave to the data, normalized to have mean value zero. With the notation

$$x_{LSr} = A_{LS} e^{i2\pi v t_r} + A_{LS}^* e^{-i2\pi v t_r}, \quad (2.8)$$

we find that the best fit to a sine wave of frequency 9.43 yr$^{-1}$ is given by

$$A_{LS} = 0.0116 - 0.0273i. \quad (2.9)$$

The corresponding fractional depth of modulation is given by $|2A_{LS}|$, which is found to be 6%. In order to evaluate the sensitivity of the Lomb-Scargle procedure, we have carried out 1,000 Monte Carlo simulations of the Super-Kamiokande 5-day data, replacing (2.7) by

$$x_{MC,r} = x_{LSr} + \sigma_r randn, \quad (2.10)$$

where $x_{LSr}$ is the expression (2.8), for the frequency $\nu = 9.43\,\text{yr}^{-1}$. (Since we are interested in the sensitivity of the Lomb-Scargle procedure to the modulation at $9.43\,\text{yr}^{-1}$, we evaluate the power at this frequency.) The result of these simulations is shown in histogram form in Figure 3. We see that there is a remarkably wide distribution of powers. Even for so small a depth of modulation as 6%, the resulting power can be as large as 20. We see that there is no simple correspondence between the depth of modulation and the resulting power in a power-spectrum analysis.

We now repeat the previous analysis, referring measurements to the mean live times rather than the mean times, as was done by Nakahata [8]. The power spectrum, computed again by the basic Lomb-Scargle method, is shown in Figure 4, and the top ten peaks are listed in Table 2. This power spectrum is consistent with that presented by Nakahata, and differs little from that shown in Figure 1.

We next carry out power-spectrum analysis using a modification of the Lomb-Scargle procedure, proposed by Scargle [17], which takes account of the experimental error estimates. Following Scargle, we introduce a weighting function

$$w_r = \frac{1/\sigma_r^2}{mean(1/\sigma_r^2)}. \qquad (2.11)$$

We then replace (2.4) by

$$S(\nu) = \frac{1}{2\sigma_0^2} \left\{ \frac{\left[\sum_r w_r x_r \cos(2\pi\nu(t_r - \tau))\right]^2}{\left[\sum_r w_r \cos^2(2\pi\nu(t_r - \tau))\right]} + \frac{\left[\sum_r w_r x_r \sin(2\pi\nu(t_r - \tau))\right]^2}{\left[\sum_r w_r \sin^2(2\pi\nu(t_r - \tau))\right]} \right\}, \qquad (2.12)$$

where $\sigma_0$ and $\tau$ are now defined by

$$\sigma_0 = std(w_r x_r) \qquad (2.13)$$

and
$$\tan(4\pi\nu\tau) = \frac{\sum_r w_r \sin(4\pi\nu t_r)}{\sum_r w_r \cos(4\pi\nu t_r)}. \qquad (2.14)$$

When we apply this procedure to the Super-Kamiokande 5-day dataset (now taking account of the mean live times, the flux estimates, and the error estimates), we obtain the power spectrum shown in Figure 5. The top ten peaks are listed in Table 3. We see that the most significant peak in this power spectrum is that at 9.43 yr$^{-1}$, with power 9.56.

The results of Monte Carlo simulations are shown in histogram form in Figure 6. We find that less than 5% of the simulations (477 out of 10,000) have power equal to or larger than the actual maximum power in the range 0 to 50 yr$^{-1}$, i.e. 9.56 at frequency 9.43 yr$^{-1}$.

## 3. LIKELIHOOD ANALYSES

In this section, we carry out power spectrum analyses using likelihood procedures [18]. Using the notation of Section 2, the log-likelihood that the data may be fit to a model that gives $X_r$ as the expected values of $x_r$ is given by

$$L = -\tfrac{1}{2} \sum_{r=1}^{R} (x_r - X_r)^2 / \sigma_r^2. \tag{3.1}$$

We estimate the power spectrum from the increase in the log-likelihood over the value expected for no modulation, corresponding to $X_r = 0$:

$$S = \tfrac{1}{2} \sum_{r=1}^{R} \frac{x_r^2}{\sigma_r^2} - \tfrac{1}{2} \sum_{r=1}^{R} \frac{(x_r - X_r)^2}{\sigma_r^2} . \tag{3.2}$$

If we assume that the data-acquisition process is uniform over the duration of each bin and examine the possibility that the flux varies sinusoidally with frequency $\nu$, the expected normalized flux estimates will be given by

$$X_r = \frac{1}{D_r} \int_{t_{sr}}^{t_{er}} dt \left( A e^{i 2\pi \nu t} + A^* e^{-i 2 \pi \nu t} \right), \tag{3.3}$$

where

$$D_r = t_{e,r} - t_{s,r} \tag{3.4}$$

and, for each frequency, the complex amplitude A is adjusted to maximize the likelihood.

The resulting power spectrum is shown in Figure 7, and the top ten peaks are listed in Table 4. The results of Monte Carlo simulations are shown in Figure 8. We find that less than 1% of the simulations (90 out of 10,000) have power equal to or larger than the actual maximum power in the range 0 to 50 yr$^{-1}$, i.e. 11.51 (for frequency 9.43 yr$^{-1}$). Hence the evidence for modulation has now increased to the 99 % significance level.

We may modify the likelihood procedure in such a way as to allow us to take account of the mean live times, as well as the start times and end times. We now replace equation (3.3) by

$$X_r = \frac{1}{D_r} \int_{t_{sr}}^{t_{er}} dt W_r(t) \left( A e^{i2\pi\nu t} + A^* e^{-i2\pi\nu t} \right), \tag{3.5}$$

where the weighting function $W_r(t)$ is chosen so that the mean value is unity,

$$\frac{1}{D_r} \int_{t_{sr}}^{t_{er}} dt W_r(t) = 1, \tag{3.6}$$

and

$$\frac{1}{D_r} \int_{t_{sr}}^{t_{er}} dt W_r(t) t = t_{ml}. \tag{3.7}$$

We seek the simplest form of the weighting function that satisfies (3.6) and (3.7). We find that these requirements are met by the following "double-boxcar" model:

$$\begin{aligned} W_r(t) &= W_{l,r} \equiv \frac{t_{e,r} - t_{ml,r}}{t_{ml,r} - t_{s,r}} \text{ for } t_{s,r} < t < t_{ml,r} \\ W_r(t) &= W_{u,r} \equiv \frac{t_{ml,r} - t_{s,r}}{t_{e,r} - t_{ml,r}} \text{ for } t_{ml,r} < t < t_{e,r} \end{aligned} \tag{3.8}$$

It appears, from studying perturbations of (3.8), that this model minimizes the difference between the maximum and minimum values of the weighting function.

We have used this modification of the likelihood method to compute the power spectrum of the Super-Kamiokande 5-day data. The result is shown in Figure 9, and the top ten peaks are listed in Table 5. The results of Monte Carlo simulations are shown in Figure 10. We again find that less than 1% of the simulations (now 74 out of 10,000) have power equal to or larger than the actual maximum power in the range 0 to 50 yr$^{-1}$, i.e. 11.67, which is found at frequency 9.43 yr$^{-1}$.

We now carry out a likelihood calculation by the "floating offset" method, in which one adjusts not only the complex amplitude for each frequency but also the offset. This method was used in our early articles on solar neutrino flux modulation [18, 19], and has also been used recently by Koshio [9]. One must be cautious in using this technique since the offset term and the sine-wave modulation term become confused at and near zero frequency, which is one reason we have not always used this method. When carrying out Monte-Carlo simulations, it is essential to exclude this region, which we do by restricting simulations to the frequency range 1 to 50 yr$^{-1}$ rather than 0 to 50 yr$^{-1}$.

Then Equation (3.2) is now replaced by

$$S = \tfrac{1}{2}\sum_{r=1}^{R} \frac{g_r^2}{s_r^2} - \tfrac{1}{2}\sum_{r=1}^{R} \frac{(g_r - G_r)^2}{s_r^2} \quad (3.9)$$

where

$$G_r = \frac{1}{D_r}\int_{t_{sr}}^{t_{er}} dt\left(C + Ae^{i2\pi\nu t} + A^* e^{-i2\pi\nu t}\right), \quad (3.10)$$

and we adjust both C and A, for each frequency, to maximize S. We show in Figure 11 the power spectrum obtained from this procedure. It is quite consistent with the power spectrum computed by Koshio [9]. The top ten peaks are listed in Table 6.

We have carried out 10,000 Monte Carlo simulations of this calculation, with results shown in Figure 12. We find that only 193 out of 10,000 simulations have power as large as or larger than the actual maximum power (11.24 at frequency 9.43 y$^{-1}$), for a significance level of 98.1%. This fraction (1.9%) is much smaller than the value (20.94%) given by Koshio on the basis of his Monte Carlo simulations. Unfortunately, there is insufficient information in Koshio's article to enable one to understand the source of this discrepancy, but since Koshio does not discuss the zero-frequency problem, it seems likely that he did not exclude the small-frequency range in carrying out his simulations.

We have calculated a probability distribution function for the modulus of the amplitude A in Equation (3.10) by evaluating the likelihood (the exponential of the log-likelihood given by Equation (3.1)) for the relevant frequency and phase. We convert this to a probability distribution function for the depth of modulation, which is related to the amplitude by

$$DOM = \frac{2|A|}{C}. \qquad (3.11)$$

This is shown in Figure 13. We see that the peak is at 6.6% and the standard deviation is 1.45%. We can be 90% confident that the amplitude is in the range 4.2% to 9.0%.

In order to evaluate the sensitivity of the likelihood procedure, taking account of a floating offset, we generate 1,000 Monte Carlo simulations of the Super-Kamiokande 5-day data, by the algorithm

$$g_{MC,r} = G_{0r} + \sigma_r randn \qquad (3.12)$$

where $G_{0r}$ is the expression (3.10) evaluated for the frequency $\nu = 9.43\,\text{yr}^{-1}$. Since we are interested in the sensitivity of the likelihood procedure to the modulation at $9.43\,\text{yr}^{-1}$, we evaluate the power at this frequency. The result of these simulations is shown in histogram form in Figure 14. We again see that there is a very wide distribution of powers. We find that 558 out of 1,000 simulations have power larger than the actual power (11.24) at frequency 9.43 yr$^{-1}$, so there is no surprise in finding a peak with power 11.24 if there is modulation with depth 6.6%.

However, one should note from Figure 14 that a search for modulation could easily run into a Type 2 error: there may be a real modulation, but the analysis may fail to reveal that fact. We find from Figure 12 that, to be 95% confident that a peak is not due to noise, the power must be 9.65 or more. However, on examining Figure 14, we find that 31% of the area of the histogram has power 9.65 or less. The implication of this comparison is the following: If there were to be a reproduction of the Super-Kamiokande experiment, and if the flux were modulated at the frequency 9.43 yr$^{-1}$ with 6.6% depth of modulation, there is a 31% chance that the experiment would *fail* to detect the modulation. On the other hand, given the power of the peaks found in the likelihood analyses, we may conclude that the probability of a Type 1 error (the inference that there is modulation when in fact there is no modulation) is in the range 1 – 2%.

4. COMPARATIVE ANALYSIS OF THE 5-DAY AND 10-DAY DATASETS

In order to understand the relationship of power spectra formed from the Super-Kamiokande 10-day and 5-day datasets, it is useful to apply the likelihood analysis to the 10-day dataset. If we use the third procedure, which allows for a floating offset, we obtain the power spectrum shown in Figure 15. The top ten peaks are listed in Table 7.

We see that the principal peak in the power spectrum is found at frequency 26.57 yr$^{-1}$, with power 11.13. The second peak is at 9.42 yr$^{-1}$ with power 7.23. These are also the two strongest peaks in the Lomb-Scargle analysis of the 10-day data carried out by Milsztajn [10]. As we have pointed out elsewhere [5, 14], the difference between the 5-day and 10-day power spectra is due primarily to aliasing. If the power spectrum of the bin-times contains a peak at frequency $\nu_T$, and if the data contains modulation at frequency $\nu_M$, then the power spectrum will also exhibit peaks at $|\nu_T - \nu_M|$ and at $\nu_T + \nu_M$. [If the peak at $\nu_T$ is particularly strong (as it is for the Super-Kamiokande datasets), the power spectrum may also exhibit peaks at $|2\nu_T - \nu_M|$ and $2\nu_T + \nu_M$, etc.] For the 10-day dataset, $\nu_T = \nu_{T10} \approx 36$ yr$^{-1}$. Since $9.42 + 26.57 = 35.99$, we may conclude that the peaks at 9.42 yr$^{-1}$ and 26.57 yr$^{-1}$ are related, one being an alias of the other. When only the 10-day dataset was available, it seemed reasonable to guess that the primary peak was that at 26.57 yr$^{-1}$ since that was the stronger of the two and could be interpreted as the second harmonic of the synodic solar rotation frequency [13]. However, analysis of the 5-day dataset [14] has made it clear that the reverse is the case: the primary peak is that at 9.42 or 9.43 yr$^{-1}$. (For the 5-day dataset, an alias peak is found at 62.56 yr$^{-1}$.) This explains why the peak at 26.57 yr$^{-1}$ (which Yoo et al. [7] refer to as a "statistical artifact") appears in the power spectrum formed from the 10-day dataset, but not in that formed from the 5-day dataset.

We may make a more objective assessment of the role of aliasing in the power spectrum formed from the 10-day dataset by using the "joint power statistic," that provides a convenient procedure for examining the correlation of two power spectra [21]. If we form the geometric mean of the powers,

$$X = (S_1 S_2)^{1/2}, \qquad (4.1)$$

the joint power statistic (of second order) is given by

$$J = -\ln(2 X K_1(2X)) \qquad (4.2)$$

where $K_1$ is the Bessel function of the second kind. This function has the following useful property: if $S_1$ and $S_2$ are distributed exponentially and are statistically independent, then J also is distributed exponentially. Hence a display of J may be interpreted in the same way as a display of a power spectrum.

Figure 16 shows the joint power statistic formed from $S(\nu)$ and $S(\nu_T - \nu)$, with $\nu_T = 35.99\ yr^{-1}$, over the frequency range 0 to 18 $yr^{-1}$. The strong peak with a nominal equivalent power 16.47 at frequency 9.43 $yr^{-1}$ shows that the peaks in the power spectrum at frequencies 9.43 $yr^{-1}$ and 26.57 $yr^{-1}$ are correlated and should be interpreted as an alias pair. (Note that we should not infer a confidence limit from the nominal equivalent power, since the two peaks are not statistically independent.)

The corresponding figure formed from the 5-day dataset is shown in Figure 17, in which the joint power statistic has been formed from $S(\nu)$ and $S(\nu_T - \nu)$, with $\nu_T = 71.99\ yr^{-1}$, over the frequency range 0 to 36 $yr^{-1}$. There is a strong peak with equivalent power 13.76 at frequency 9.43 $yr^{-1}$, which is formed from the peak at 9.43 $yr^{-1}$ and from a peak at 62.56 $yr^{-1}$.
We see that aliasing plays a lesser role in the power-spectrum analysis of the 50-day dataset, since the timing frequency is much higher (72 $yr^{-1}$ instead of 36 $yr^{-1}$). We note also that the peak at 12.31 $yr^{-1}$ appears in both Figure 16 and Figure 17, showing that it is accompanied by aliases in both the 10-day and 5-day datasets. This peak, if real, cannot be attributed to rotational modulation in the convection zone, but the uncertainties in the rotation rate in the radiative zone do not preclude its attribution to rotational modulation in that region.

## 5. DISCUSSION

Although we have presented the analyses of Sections 2 and 3 in their conventional forms, and introduced the likelihood procedure of Sections 4, 5 and 6 as something different, we may in fact regard all the analyses presented in this article as special cases of the likelihood procedure. The relationship is shown schematically in Figure 18. Panels (a) and (b) show the "single boxcar" and "double boxcar" weighting functions corresponding to the uniform weighting in Equation (3.3) and the non-uniform weighting in Equation (3.5). If one calculates the power spectrum from the likelihood procedure, adopting the standard deviation of the flux estimates as the error term and using a delta-function form of the time weighting function, as in panels (c) and (d), one retrieves the power spectra computed by the Lomb-Scargle procedure in Section 2.. The third calculation of Section 2 is equivalent to using the time weighting function shown in panel (d) and the actual error estimates.

We now compare the results of these analyses with those of previous publications. Milsztajn [10] used the basic Lomb-Scargle method to analyze the 10-day dataset, assigning flux measurements to the mean times. Hence his method was the first presented in Section 2. Milsztajn's power spectrum is similar to that obtained in Section 4 (see Figure 15), showing two principal peaks at frequencies 26.57 $yr^{-1}$ and 9.42 $yr^{-1}$. In his article, Milsztajn states "…the sampling, though quite regular, is sufficiently variable that no aliasing is observed…" However, in fact there is aliasing: we saw in Section 4 that the two principal peaks comprise an alias pair, related by the timing frequency 35.99 $yr^{-1}$.

Nakahata [8] also carried out a Lomb-Scargle analysis of the 10-day dataset, assigning measurements to the mean times, and obtained a power spectrum close to those found previously [10, 13]. However, Nakahata had access to the mean live time measurements, and therefore repeated the Lomb-Scargle analysis, assigning flux measurements to the mean live times rather than to the mean times. This analysis yields a peak at frequency 26.55 yr$^{-1}$ with power 7.51, and a peak at frequency 9.42 yr$^{-1}$ with power 6.67. Nakahata interprets the second peak as "a natural peak in the random distribution" (by which we presume he means a statistical fluctuation) whereas, as we have seen in Section 4, the peaks at 26.55 yr$^{-1}$ and 9.42 yr$^{-1}$ are an alias pair.

Yoo et al. [7] were the first to have access to and analyze the 5-day dataset. Their analysis is that reproduced in Section 3, leading to the power spectrum shown in Figure 4. Yoo et al. commented on our analysis of the 10-day dataset [13] and asserted that the difference in the resulting power spectra was due to the fact that our analysis used the mean times rather than the mean live times, but this statement was incorrect, since our analysis used the start times and end times, and made no reference to the mean times. Yoo et al. noted that the peak at 26.55 yr$^{-1}$, which was evident in the 10-day power spectrum, was no longer evident in the 5-day power spectrum, and concluded that this "provides additional confirmation that the [peak at 26.55 yr$^{-1}$] in the 10-day long sample is a statistical artifact." However, as we have seen, the peak disappeared because it is an alias of the peak at 9.43 yr$^{-1}$ in the 10-day power spectrum but not in the 5-day power spectrum.

Since Yoo et al. [7] concluded that their power spectrum analysis did not yield evidence for periodic modulation of the solar neutrino flux, they included in their article a "sensitivity" calculation designed to determine the significance of this null result. It was designed to answer the following question: If there were a modulation of specified amplitude at specified frequency, what is the probability that this would have resulted in a positive outcome in their power-spectrum analysis? [Since we did not catch any fish, we had better examine our nets.] Their procedure comprised a set of Monte Carlo calculations in which fictitious flux estimates are generated by (a) computing the flux estimates to be expected from neutrino flux with specified sinusoidal modulation; (b) adding a Gaussian random fluctuation with width determined by the actual error estimates; and (c) calculating the resulting Lomb-Scargle power spectrum. They found that, for periods of 20 days or more (frequencies of 18 yr$^{-1}$ or less), this method could not reliably (i.e. with 95% probability) identify the signal if the depth of modulation is less than 10%. There is no conflict between their conclusion and the result of our Lomb-Scargle analyses presented in Section 2, since we found that led to an estimate of the modulation at 9.43 yr$^{-1}$ (period 38.73 days) of only 6%.

In our analyses, we have found evidence for periodic modulation, so our motivation for Monte-Carlo simulations is very different from that of Yoo et al. Our goal is not to understand what we might have missed, but to check on what we have found. [If you do not catch a fish, you examine the net; if you do catch a fish, you examine the fish.] We found (Figure 3) that we were in a sense a little unlucky in our basic Lomb-Scargle analyses (that take no account of the error estimates) for the following reason: if there were 1,000 experiments identical in design and operation to the Super-Kamiokande experiment, and if the solar neutrino flux were modulated at the frequency 9.43 yr$^{-1}$ with depth of modulation 7%, then 650 of the experiments would have yielded a power at that frequency that is larger than that actually found in the Super-Kamiokande dataset. Indeed, a few percent of those experiments would have yielded a power of 15 or more.

On the other hand, we found in our likelihood analyses (Figure 14) that it is perfectly reasonable that we found a positive outcome due to the same assumed modulation. A comparison of Figures 3 and 14 shows that the likelihood method (that takes account of the start time, end time, the flux estimate and error estimates) is a more sensitive detector of modulation than the basic Lomb-Scargle method that takes account only of the flux estimate and one item of timing information (mean time or mean live time). This result is not unreasonable. In investigating a hypothesis, it makes sense to process as much information as possible. (The Super-Kamiokande collaboration produced and analyzed the 5-day dataset because it contains more information than the 10-day dataset.)

Koshio [9] has published an analysis from which he concludes that there is no evidence for periodic modulation in the Super-Kamiokande 5–day dataset. We found in Section 3 that we could reproduce his power spectrum but not the results of his Monte-Carlo simulations. We speculated that Koshio may have examined simulations over a frequency band extending to zero frequency, which would result in anomalously large powers in some fraction of the simulations. In simulations that allow for a "floating offset," it is crucial that one avoids frequencies at or near zero.

We now comment further on a point raised in Section 2. The Monte Carlo analysis of Nakahata [8], Yoo [7], and Koshio [9] implicitly assumes that, a priori, all frequencies in the chosen search band (here 0 to 50 $yr^{-1}$) are equally likely. This may be appropriate if one has no idea what mechanism might lead to periodic modulation of the solar neutrino flux. However, if one were considering (for instance) the specific possibility that the solar neutrino flux may exhibit a periodic modulation due to solar rotation, then the appropriate search band would be determined by the Sun's internal quasi-equatorial synodic rotation rate. For modulations in the convection zone, this points to synodic frequencies in the range 13.4 to 13.8 $yr^{-1}$. For modulations in the tachocline, the range is 12.8 to 13.4 $yr^{-1}$. For modulations in the radiative zone, there is great uncertainty concerning the appropriate limits. Analysis of MDI helioseismology data [22] yields a one-sigma frequency range of 10.3 to 14.5 $yr^{-1}$, and a two-sigma range of 8.2 to 16.6 $yr^{-1}$. Hence one cannot at this stage rule out the possibility that the prominent modulation with frequency 9.43 $yr^{-1}$ may prove to be due to rotation in the deep interior of the Sun.

In searching for rotational modulation, it makes sense to examine not only the known range of rotation rates of the solar interior, but also multiples (harmonics) of this range. The peak at 39.28 $yr^{-1}$ may be attributed to the "third" harmonic (3 times the fundamental frequency) of the rotation frequency in the same region, which is found to be much more prominent than the fundamental and the "second" harmonic in an analysis of the disk-center solar magnetic field at that time [5, 14]. If one were looking for other types of modulation, such as r-modes [23 - 25], it makes sense to determine search bands appropriate for those modulations. In our analysis of the 5-day dataset [5, 14] that uses the procedure summarized in Section 4, we point out that the two strongest peaks (at 9.43 $yr^{-1}$ and 43.72 $yr^{-1}$) may both be due to an internal r-mode oscillation with indices l = 2, m = 2, occurring where the sidereal rotation rate is about 14.15 $yr^{-1}$, which would place it inside or just above the tachocline. It should be noted that r-modes are excited individually. For instance, one may find evidence of the l = 3, m =2 r-mode (period 77 days) or the l = 3, m = 3 r-mode (period 52 days) without there being any concurrent evidence of the l = 3, m = 1 (154 day) oscillation [25, 26, 27].

There is another basic point that is worth noting. Power spectrum analysis of solar neutrino data may detect an oscillation in the neutrino flux if the flux is modulated by a stable, high-Q oscillation. However, it is possible that the flux is variable, but the variability does not meet these criteria, in which case the variability may well escape detection by power-spectrum analysis. For instance, power-spectrum analysis is not well suited to the detection of a stochastic variation, and it may fail to detect an oscillation that drifts in frequency and/or jumps in phase. Hence even if a power spectrum analysis were to fail to reveal a peak in a wide frequency range (which it does not), this in itself would not comprise evidence that the flux is constant. These considerations invalidate the following assertion by Yoo et al. [7]: " Based on the observation of no significant periodicity, SK-I data exclude modulations greater than 10% of the $^8$B neutrino flux arising as a result of more than 0.4% changes in the solar core temperature, allowing a new measure of the solar core's stability."

The above re-analysis of Super-Kamiokande data supports our earlier conclusion that there is evidence for an intrinsic variability of the solar neutrino flux [5, 14], probably originating at or near the tachocline. In principle, such a modulation could be due to the RSFP (Resonant-Spin-Flavor-Precession) process [2]. Such an RSFP effect involving only the three active neutrinos would occur in the solar core, followed by an MSW transition at a larger radius [4]; this would be incompatible with the inference that modulation occurs at the tachocline or above. In this context, we may also note that our analysis of GALLEX data [5, 28, 29, 30] gives evidence for modulation at 13.59 yr$^{-1}$ which, if interpreted as a synodic rotation frequency, corresponds to a sidereal rotation frequency of 14.59 yr$^{-1}$, placing the process in the convection zone [28, 29]. Also we observed [5] that the modulation changes at a change in the solar cycle, which is compatible with the assumption that the effect occurs in the convection zone, and not obviously compatible with the assumption that it occurs in the stable radiative zone. Because of this change of convection-zone field with solar cycle, even the closely related GALLEX (solar cycle 23) and GNO (solar cycle 24) experiments do not show the same frequencies [30]. GALLEX data show evidence of modulation at the fundamental and "second" harmonic (twice the fundamental frequency) of a frequency in the rotation band, whereas Super-Kamiokande data show evidence of modulation at the "third" harmonic (three times the fundamental frequency). We note that GNO, which shares a solar cycle with Super-Kamiokande, shows evidence of the l=2, m=2 r-mode oscillation which is prominent in the Super-Kamiokande power spectrum, even though the two experiments measure very different neutrino energy ranges.

We also note that an RSFP process in the radiative zone leads to a much smaller depth of modulation (about 2% [4]) than that (about 7%) which we find from power-spectrum analysis. On the other hand, a 7% depth of modulation is in the range predicted [6] for the model [5] in which a sterile neutrino couples to active neutrinos only via a transition magnetic moment. This model is compatible with known limitations on sterile neutrinos, and with the present null measurements of solar antineutrinos. In this model, for which the sterile neutrino and the electron neutrino have a very small mass difference, the RSFP process occurs in the solar convection zone at a larger radius than the location of the MSW effect. It appears that current evidence for variability of the solar neutrino flux from the Homestake (99.9% CL) [13, 14], GALLEX (99.9% CL) [28, 29, 30], and Super-Kamiokande (98 - 99% CL) [13, 14] experiments, when taken together, comprise a strong case for the existence of a sterile neutrino which, together with the needed large transition magnetic moment, would be harbingers of important new physics.

Clearly, it is necessary to pursue further the issue of variability of the solar neutrino flux. It appears (not unreasonably) that the most sensitive methods are those that process the greatest amount of relevant information. It is also clearly desirable to package data into bins shorter than 5 days. Both goals could be met by packaging data into 1-day bins. However, with such short bins, it may not be adequate to summarize data for each bin in terms of a most-likely flux and upper and lower error estimates. It may instead be necessary to summarize the data in terms of a probability distribution function for the flux. It would obviously be most helpful if the SNO collaboration were to make their data publicly available, and it would be especially helpful if the Super-Kamiokande and SNO collaborations would provide their data in identical 1-day bins, tied perhaps to Universal Time.

Research reported in this article was supported in part (for PAS) by NSF grant AST-0097128.

FIGURES

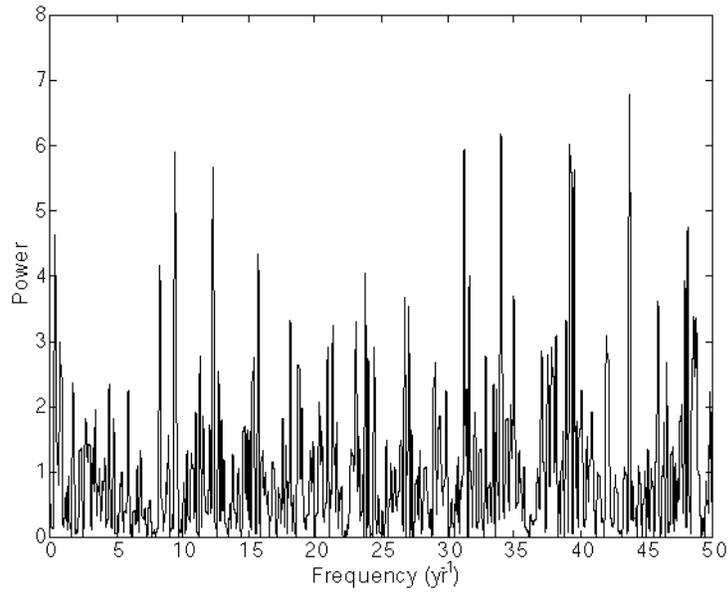

Figure 1. Power spectrum of 5-day Super-Kamiokande data formed by the basic Lomb-Scargle method, using the mean times.

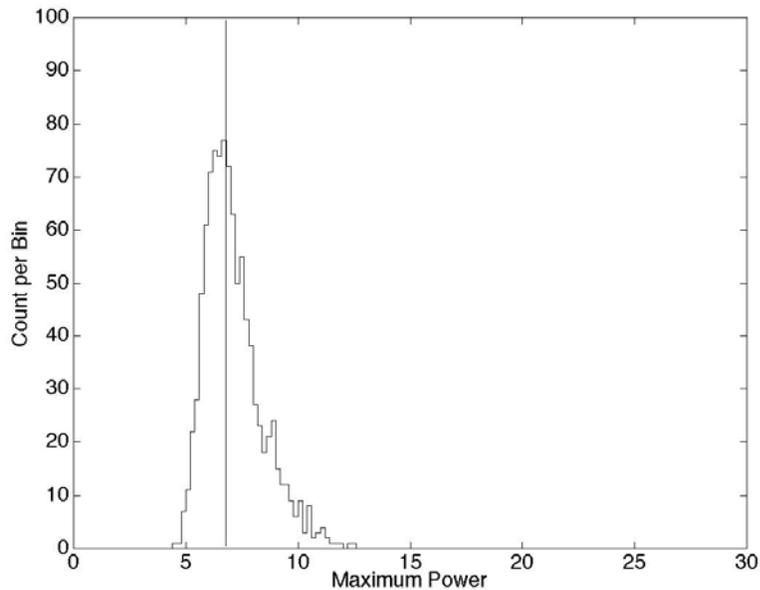

Figure 2. Histogram display of the maximum power, computed by the Lomb-Scargle procedure using the mean times, over the frequency band 0 to 50 yr$^{-1}$, for 1,000 Monte Carlo simulations of the Super-Kamiokande 5-day data. 485 out of 1,000 simulations have power larger than the actual maximum power (6.79 at frequency 43.72 yr$^{-1}$). 824 out of 1,000 simulations have power larger than the power (5.90) at frequency 9.43 yr$^{-1}$.

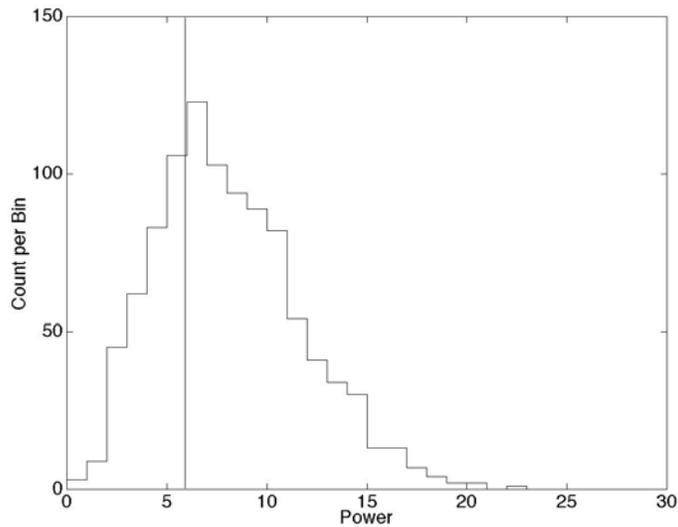

Figure 3. Histogram display of the power, computed by the Lomb-Scargle procedure using the mean times, at the frequency 9.43 yr$^{-1}$, for 1,000 Monte Carlo simulations of the Super-Kamiokande 5-day data. Each simulation contains the actual sine-wave term at 9.43 yr$^{-1}$, plus normally distributed random terms. 650 out of 1,000 simulations have power larger than the power (5.90) at frequency 9.43 yr$^{-1}$.

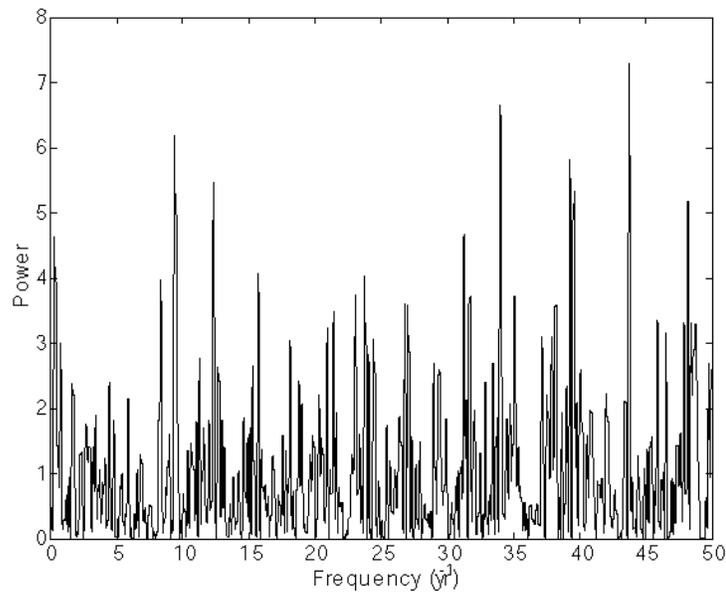

Figure 4. Power spectrum of 5-day Super-Kamiokande data formed by the basic Lomb-Scargle method, using the mean live times.

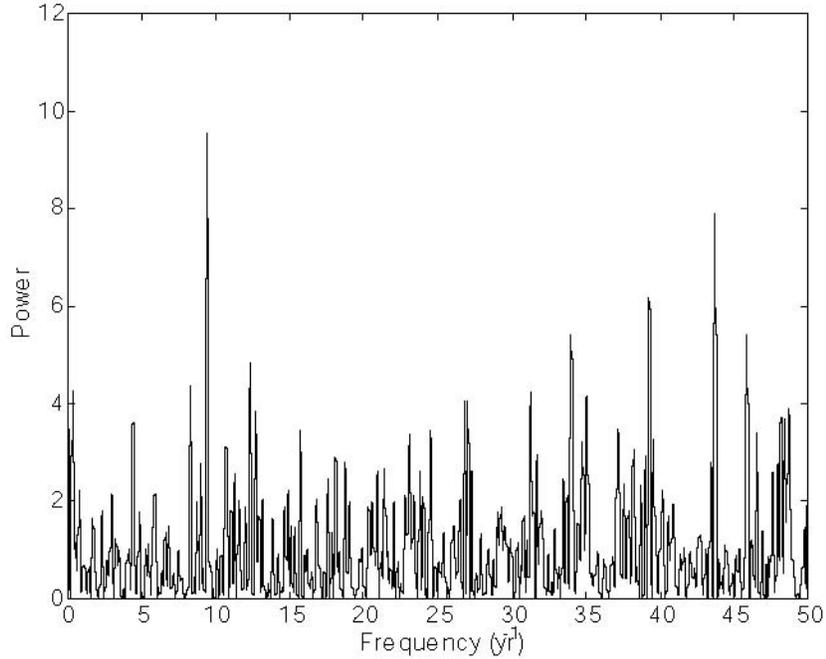

Figure 5. Power spectrum of 5-day Super-Kamiokande data, using the mean live times, formed by the modified Lomb-Scargle method.

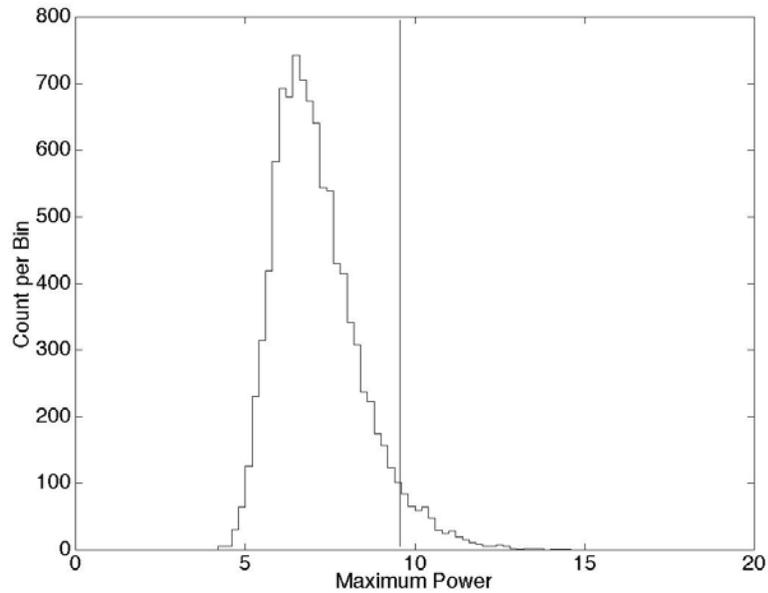

Figure 6. Histogram display of the maximum power, computed by the modified Lomb-Scargle procedure using the mean live times, over the frequency band 0 to 50 yr$^{-1}$, for 10,000 Monte Carlo simulations of the Super-Kamiokande 5-day data. 477 out of 10,000 simulations have power larger than the actual maximum power (9.56 at frequency 9.43 yr$^{-1}$).

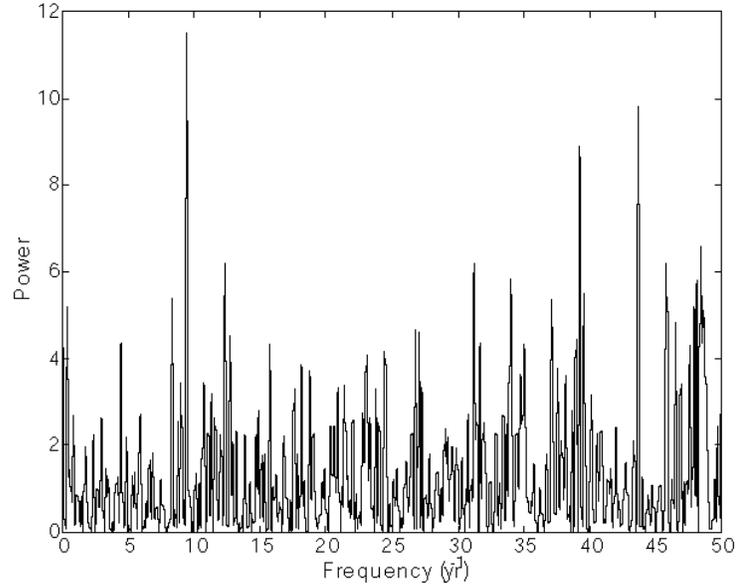

Figure 7. Power spectrum of 5-day Super-Kamiokande data, formed by the likelihood method, using the start and end times.

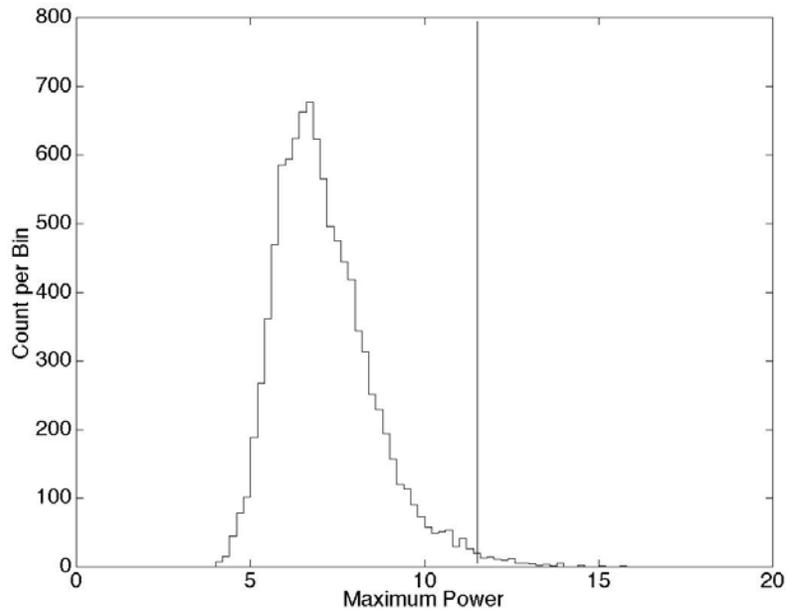

Figure 8. Histogram display of the maximum power, formed by the likelihood method, using the start and end times, over the frequency band 0 to 50 $yr^{-1}$, for 10,000 Monte Carlo simulations of the Super-Kamiokande 5-day data. 90 out of 10,000 simulations have power larger than the actual maximum power (11.51 at frequency 9.43 $yr^{-1}$).

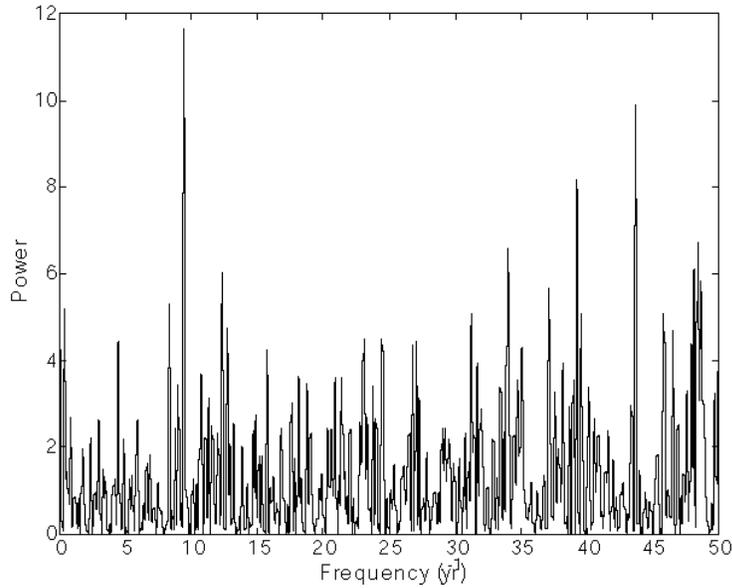

Figure 9. Power spectrum of 5-day Super-Kamiokande data, using the start times, end times, and mean live times, formed by the likelihood method.

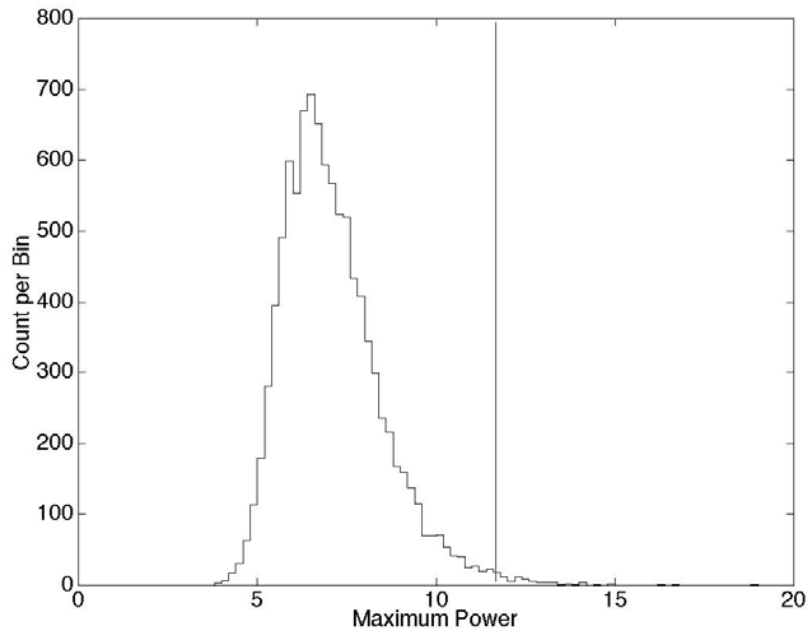

Figure 10. Histogram display of the maximum power, computed by the likelihood method using the start times, end times, and mean live times, over the frequency band 0 to 50 $yr^{-1}$, for 10,000 Monte Carlo simulations of the Super-Kamiokande 5-day data. 74 out of 10,000 simulations have power larger than the actual maximum power (11.67 at frequency 9.43 $yr^{-1}$).

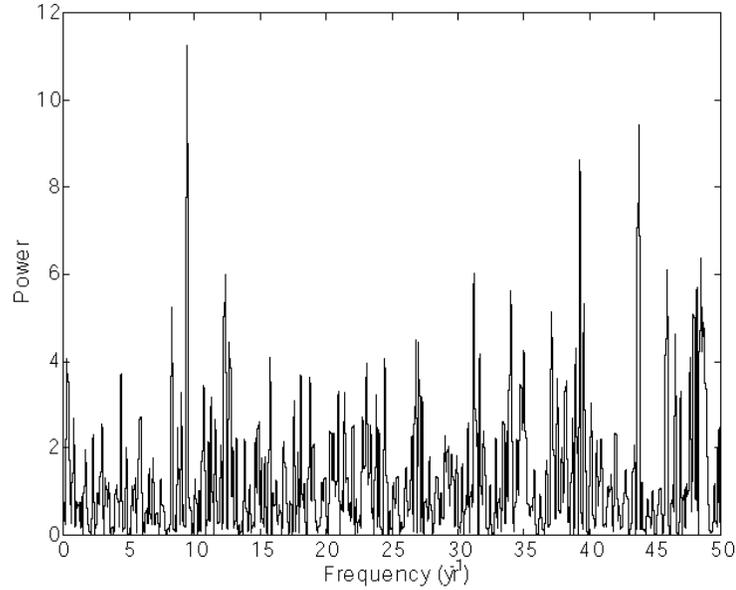

Figure 11. Power spectrum of 5-day Super-Kamiokande data, using the start times and end times, and allowing for a floating offset, formed by the modified SWW likelihood method.

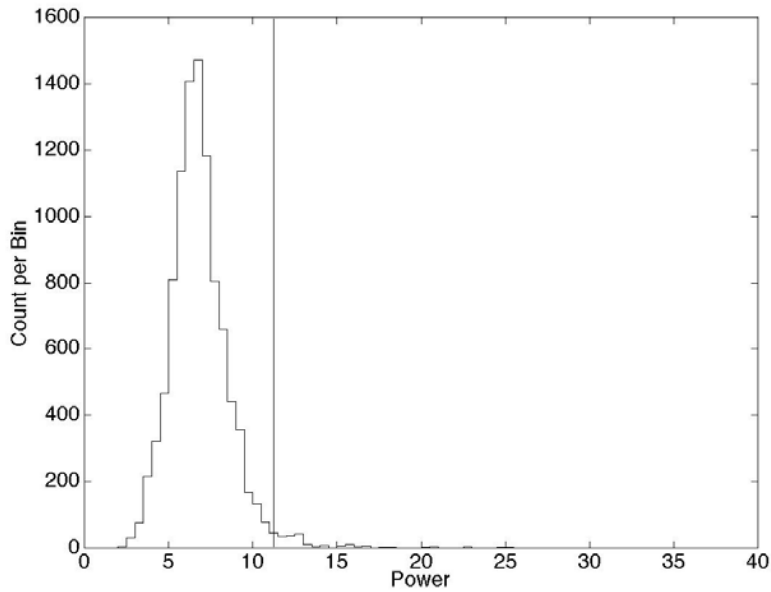

Figure 12. Histogram display of the maximum power, computed by the likelihood method using the start times and end times and allowing for a floating offset, over the frequency band 1 to 50 $yr^{-1}$, for 10,000 Monte Carlo simulations of the Super-Kamiokande 5-day data. 193 out of 10,000 simulations have power larger than the actual maximum power (11.24 at frequency 9.43 $yr^{-1}$).

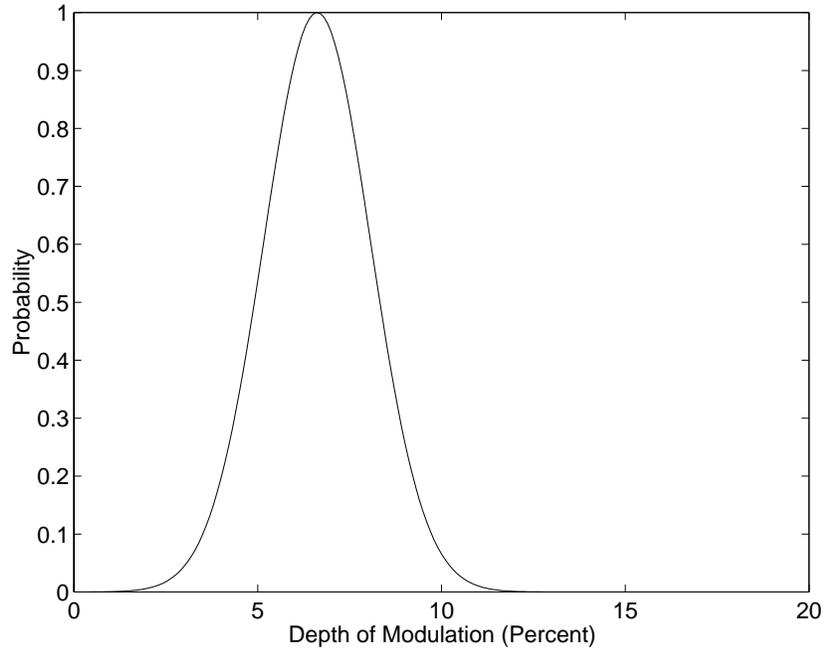

Figure 13. Probability distribution function for the depth of modulation at 9.43 yr-1. We see that the peak is at 6.6%, the standard deviation is 1.45%, and there is 90% probability that the depth of modulation is in the range 4.2% to 9.0%.

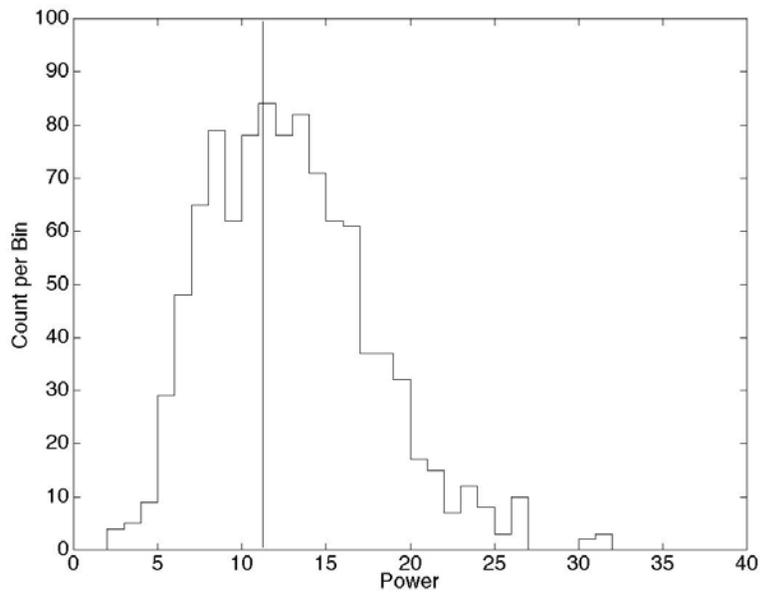

Figure 14. Histogram display of the power at 9.43 yr$^{-1}$, computed by the floating-offset likelihood procedure, for 1,000 Monte Carlo simulations of the Super-Kamiokande 5-day data. Each simulation contains the actual sine-wave term at 9.43 yr$^{-1}$, plus normally distributed random terms. 558 out of 1,000 simulations have power larger than the power (11.24) at frequency 9.43 yr$^{-1}$.

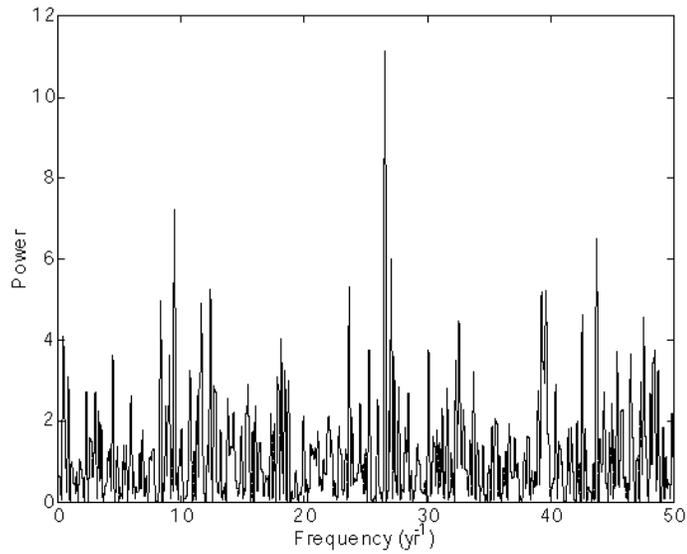

Figure 15. Power spectrum of 10-day Super-Kamiokande data, using the start times and end times, formed by a likelihood method that allows for a floating offset.

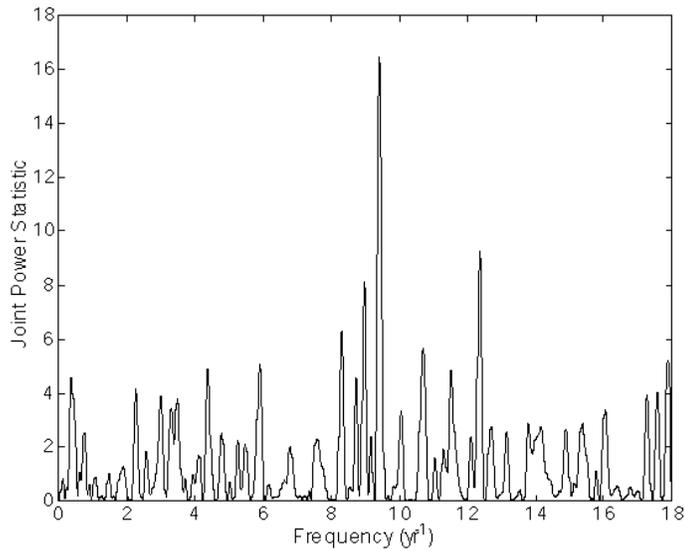

Figure 16. Joint spectrum statistic formed from the power spectrum formed from the 10-day Super-Kamiokande data by combining the power at frequency $\nu$ with that at frequency $\nu_T - \nu$ where $\nu_T \ (\approx 36\,\text{yr}^{-1})$ is the timing frequency.

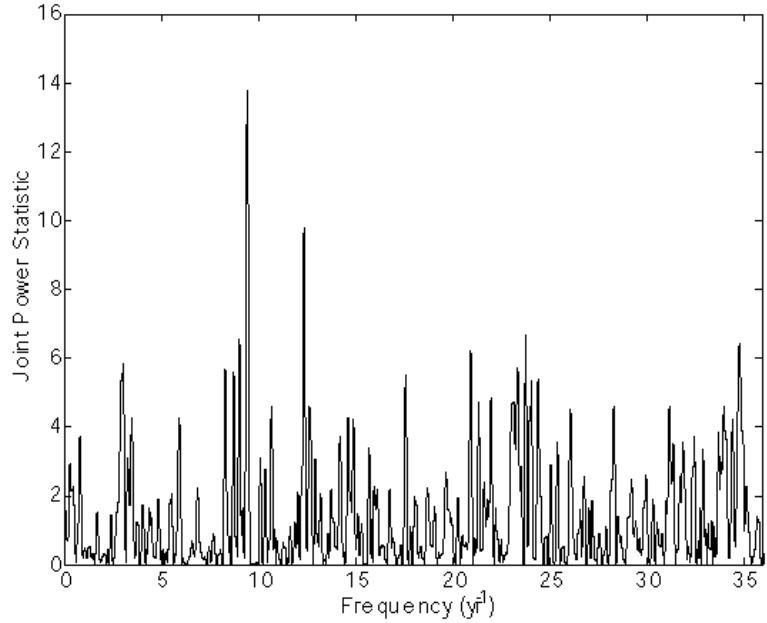

Figure 17. Joint spectrum statistic formed from the power spectrum formed from the 5-day Super-Kamiokande data by combining the power at frequency $\nu$ with that at frequency $\nu_T - \nu$ where $\nu_T \left( \approx 72 \, \text{yr}^{-1} \right)$ is the timing frequency.

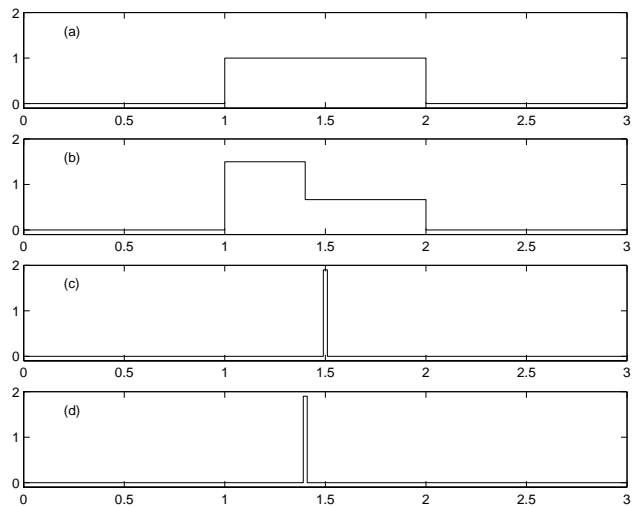

Figure 18. (a) Time window function for uniform weight over start time to end time. (b) Time window function with non-uniform weight to take account of mean live time. (c) Delta-function form for time window function at mid-point of bin. (d) Delta-function form for time window function at mean live time.

TABLE 1

Top ten peaks in the basic Lomb-Scargle power spectrum,
using the mean times of bins.

| Order | Frequency (yr$^{-1}$) | Power |
|---|---|---|
| 1 | 43.72 | 6.79 |
| 2 | 34.02 | 6.19 |
| 3 | 39.28 | 6.03 |
| 4 | 31.23 | 5.95 |
| 5 | 9.43 | 5.90 |
| 6 | 12.31 | 5.67 |
| 7 | 39.54 | 5.65 |
| 8 | 48.16 | 4.75 |
| 9 | 0.36 | 4.64 |
| 10 | 15.73 | 4.35 |

TABLE 2

Top ten peaks in the basic Lomb-Scargle power spectrum,
using the mean live times of bins.

| Order | Frequency (yr$^{-1}$) | Power |
|---|---|---|
| 1 | 43.73 | 7.29 |
| 2 | 34.01 | 6.65 |
| 3 | 9.43 | 6.18 |
| 4 | 39.27 | 5.82 |
| 5 | 12.31 | 5.48 |
| 6 | 39.54 | 5.34 |
| 7 | 48.15 | 5.18 |
| 8 | 31.23 | 4.67 |
| 9 | 0.36 | 4.64 |
| 10 | 15.73 | 4.06 |

TABLE 3

Top ten peaks in the modified Lomb-Scargle power spectrum.

| Order | Frequency (yr$^{-1}$) | Power |
|---|---|---|
| 1 | 9.43 | 9.56 |
| 2 | 43.72 | 7.91 |
| 3 | 39.28 | 6.18 |
| 4 | 33.99 | 5.42 |
| 5 | 45.85 | 5.42 |
| 6 | 12.31 | 4.86 |
| 7 | 8.30 | 4.38 |
| 8 | 0.34 | 4.26 |
| 9 | 31.25 | 4.23 |
| 10 | 35.04 | 4.15 |

TABLE 4

Top ten peaks in a likelihood power spectrum,

Using start and end times.

| Order | Frequency (yr$^{-1}$) | Power |
|---|---|---|
| 1 | 9.43 | 11.51 |
| 2 | 43.72 | 9.83 |
| 3 | 39.28 | 8.91 |
| 4 | 48.43 | 6.57 |
| 5 | 12.31 | 6.21 |
| 6 | 31.24 | 6.20 |
| 7 | 45.86 | 6.20 |
| 8 | 34.00 | 5.83 |
| 9 | 48.16 | 5.78 |
| 10 | 39.55 | 5.49 |

TABLE 5

Top ten peaks in a likelihood power spectrum,

Using start times, end times, and mean live times.

| Order | Frequency (yr$^{-1}$) | Power |
|---|---|---|
| 1 | 9.43 | 11.67 |
| 2 | 43.72 | 9.87 |
| 3 | 39.28 | 8.18 |
| 4 | 48.43 | 6.72 |
| 5 | 33.99 | 6.58 |
| 6 | 48.16 | 6.09 |
| 7 | 12.31 | 6.05 |
| 8 | 48.69 | 5.84 |
| 9 | 37.12 | 5.65 |
| 10 | 8.30 | 5.32 |

TABLE 6

Top ten peaks in a likelihood power spectrum computed from the 5-day dataset,

Using start times, end times, and allowing for a floating offset.

| Order | Frequency (yr$^{-1}$) | Power |
|---|---|---|
| 1 | 9.43 | 11.24 |
| 2 | 43.72 | 9.44 |
| 3 | 39.28 | 8.64 |
| 4 | 48.43 | 6.38 |
| 5 | 45.86 | 6.10 |
| 6 | 31.24 | 6.03 |
| 7 | 12.31 | 6.01 |
| 8 | 48.16 | 5.69 |
| 9 | 33.99 | 5.63 |
| 10 | 39.55 | 5.32 |

TABLE 7

Top ten peaks in a likelihood power spectrum computed from the 10-day dataset,
Using start times and end times, and allowing for a floating offset.

| Order | Frequency (yr$^{-1}$) | Power |
|---|---:|---:|
| 1 | 26.57 | 11.13 |
| 2 | 9.42 | 7.23 |
| 3 | 43.73 | 6.52 |
| 4 | 27.02 | 6.00 |
| 5 | 23.63 | 5.32 |
| 6 | 12.36 | 5.26 |
| 7 | 39.59 | 5.22 |
| 8 | 39.31 | 5.19 |
| 9 | 8.31 | 4.99 |
| 10 | 11.56 | 4.92 |